\documentclass[conference]{IEEEtran}

\usepackage[utf8]{inputenc}
\usepackage[T1]{fontenc}
\usepackage{cite}
\usepackage[pdftex]{graphicx}
\usepackage{amsmath}
\usepackage{bm}
\usepackage{amssymb}
\usepackage{algorithmic}
\usepackage{array}
\usepackage[caption=false,font=normalsize,labelfont=sf,textfont=sf]{subfig}
\usepackage{stfloats}
\usepackage{url}
\usepackage{enumerate}	
\usepackage[dvipsnames]{xcolor}
\usepackage{amsthm}
\usepackage[normalem]{ulem}
\usepackage{cancel}
\usepackage{verbatim}

\theoremstyle{remark}
\newtheorem{remark}{Remark}
\newtheorem{assumption}{Assumption}

\begin{document}

\title{A selection of PID type controller settings via LQR approach for two-wheeled balancing robot}

\author{\IEEEauthorblockN{Krzysztof Laddach, Mateusz Czyżniewski, Rafa\l{} \L{}angowski}
\IEEEauthorblockA{Department of Electrical Engineering, Control Systems and Informatics, Gda\'nsk University of Technology\\
ul. G. Narutowicza 11/12, 80-233 Gda\'nsk, Poland\\
Email: krzysztof.laddach@pg.edu.pl, mateusz.czyzniewski@pg.edu.pl, rafal.langowski@pg.edu.pl}}

\maketitle

\begin{abstract}
The problem of PID type controller tuning has been addressed in this paper. In particular, a method of selection of PD settings based on the solution of linear--quadratic optimisation problem using the energy criterion has been investigated. Thus, the possibility of transforming optimal settings of the linear--quadratic regulator into the settings of the controller in the classical control system has been given. The presented methodology has been used during synthesis of control system for a two--wheeled balancing robot. Finally, the performance of the proposed control system has been validated by simulation in Matlab/Simulink environment with the use of a two--wheeled balancing robot model.
\end{abstract}


\IEEEpeerreviewmaketitle

\section{Introduction} \label{sec:introduction}

A control problem of a non--linear and unstable plant is one of the important challenges in control engineering. A lot of mechanical dynamic systems are the example of such plants. Moreover, these plants often are under--actuated, i.e. they have more controlled variables than the number of control inputs. Hence, from this point of view they belong to SIMO (single input multiple output) systems. This class of systems is  widely used  for  modelling  plants like, e.g., different kind of pendulums, e.g., \cite{Furuta:1991,Jadlovska:2012,Andrzejewski:2019,Waszak:2020} and balancing robots and vehicles, e.g., \cite{Nguyen:2004,Dai:2015,Velazquez:2016}. It is known that the one of main aims of control of these systems is their stabilisation at a given equilibrium point. Due to, i.a., the non--linear dynamics of considered plant, it is possible to distinguish two main approach to stabilisation (control) problem. The first approach is based on non--linear control algorithms, e.g., sliding mode control and fuzzy logic controller \cite{Park:2009,Romlay:2019}. Whereas the second approach uses linear controllers based on a negative feedback loop from controlled or state variables \cite{Wang:2011,Prasad:2011,Mahapatra:2017,Waszak:2020}. A widespread approach using controlled variables feedback loop is based on PID type controllers. Therefore, different kind of methods are addressed to selection of PID controller settings. Starting from experimental solutions, through analytical and engineering (based on concepts such as Ziegler-Nichols) selection of controller settings to methods based on solving optimisation tasks \cite{Astrom:1995}.

\begin{figure}[!b]
\centering
\includegraphics[width=1.7in]{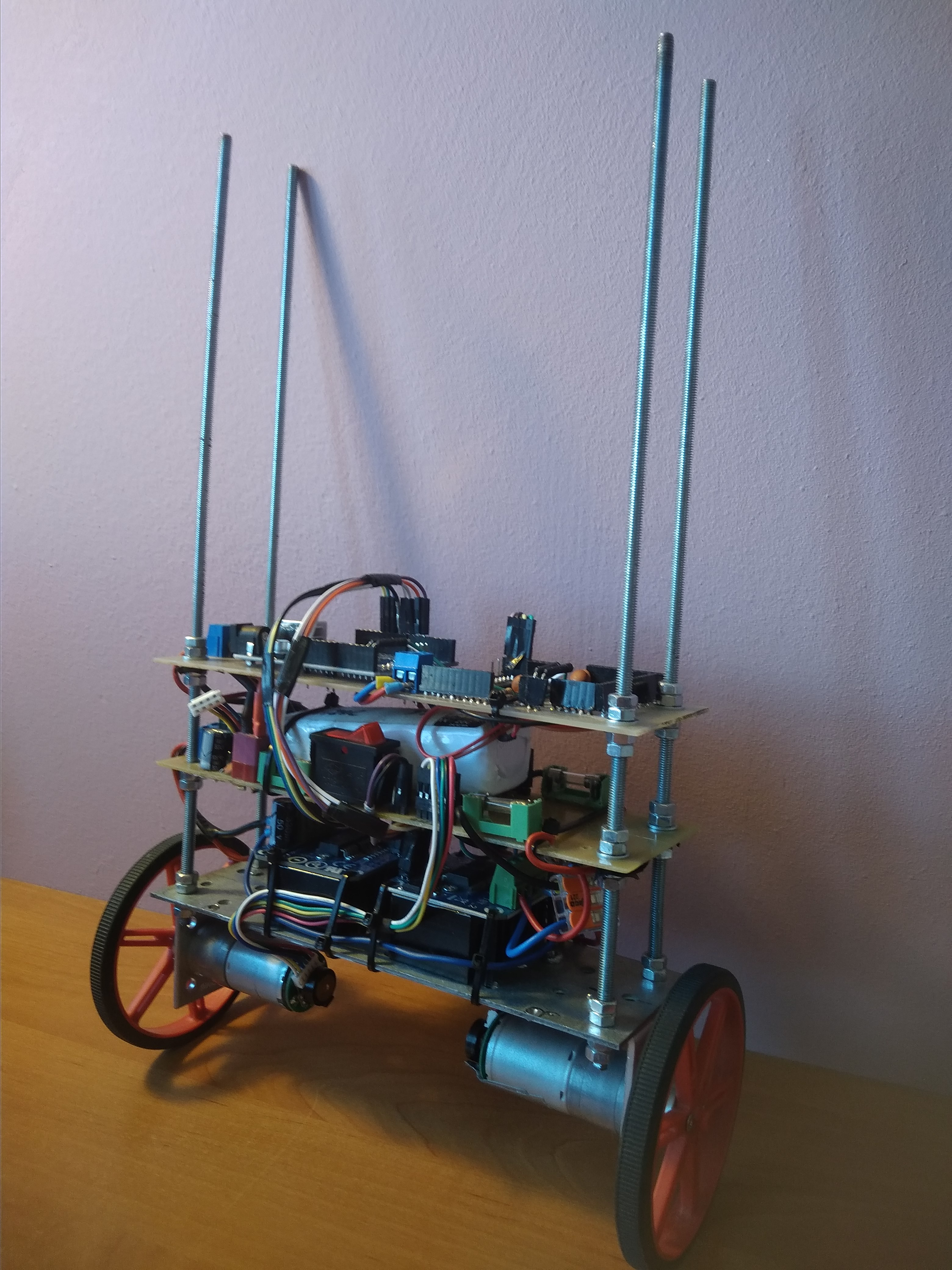}
\caption{The considered two-wheeled balancing robot.}
\label{fig:robot}
\end{figure}

The goal of this work is to provide an alternative  method of PID tuning, that uses optimum setting values (gains) of the controller based on the state feedback \cite{Astrom:2008}. Clearly, the selection of PID type controller settings is based on the solution of the linear--quadratic optimisation task. Such an approach can be found in the literature, primarily for given applications of SISO (single input single output) plants, e.g., \cite{He:2000}. Whereas, in this paper the general considerations allow using this technique for a certain class of SIMO plants are presented. The presented work is an extension of the considerations contained in \cite{Czyzniewski:2019}. Firstly, an alternative approach to model linearisation has been used. Secondly, the devised methodology has been used for control system design purposes of a two-wheeled balancing robot. In this paper, the considered robot can be described as a mobile (passive) single-arm inverted pendulum and it is shown in Fig. \ref{fig:robot}. The derived control systems with PD controllers as well as linear--quadratic regulator (LQR) have been implemented and validated in Matlab/Simulink environment using a balancing robot model \cite{Laddach:2019}.



\section{Problem formulation} \label{sec:problem_form}

The general class of considered SIMO systems represents mechanical dynamic systems which can be modelled using Newtonian (classical), Lagrangian or Hamiltonian mechanics and it can be given as follows \cite{Slotine:1991}:
\begin{equation}\label{eq:general_form_of_mechanical_model}
    \begin{cases}
        \ddot{\bm{\Theta}}(t) &= - \bm{C}(\dot{\bm{\Theta}}(t),\bm{\Theta}(t))\dot{\bm{\Theta}}(t) - \bm{G}(\bm{\Theta}(t))\\ &+ \bm{Q}(\bm{\Theta}(t))u(t),\\
        \dot{\bm{\Theta}}(t_0) &= \dot{\bm{\Theta}}_0 \\ \bm{\Theta}(t_0) &= \bm{\Theta}_0,
    \end{cases}
\end{equation}
where: $\dot{(\cdot)}$ and $\ddot{(\cdot)}$ are the first and second derivative with respect to $t$, respectively; $t \in \mathbb{R}_{+} \cup \{ 0 \}$ is time, $\mathbb{R}$, $\mathbb{R}_{+}$ are a real number field and its positive part, respectively; $\bm{\Theta}(t) \in \mathbb{R}^{q}$ denotes a vector of linear and angular displacements, where $q \in \mathbb{N}_+$ is a number of controlled outputs and $\mathbb{N}_+$ is a positive part of a natural number set; $u(t) \in \mathbb{R}$ denotes a control input; $\bm{C}(\dot{\bm{\Theta}}(t),\bm{\Theta}(t)) = \begin{bmatrix} \bm{C}_1(\cdot)^{\mathrm{T}} \ldots \bm{C}_q(\cdot)^{\mathrm{T}} \end{bmatrix}^{\mathrm{T}}$, $\bm{G}(\bm{\Theta}(t)) = \begin{bmatrix} G_1(\cdot) \ldots G_q(\cdot) \end{bmatrix}^{\mathrm{T}}$ and $\bm{Q}(\bm{\Theta}(t)) = \begin{bmatrix} Q_1(\cdot) \ldots  Q_q(\cdot) \end{bmatrix}^{\mathrm{T}}$ signify non-linear matrix and vector value smooth functions which arguments are $\bm{\Theta}(t)$ and $\dot{\bm{\Theta}}(t)$; $\bm{\Theta}_0$, $\dot{\bm{\Theta}}_0$ denote initial conditions.

It should be added that, from a physical construction point of view, the systems determined through \eqref{eq:general_form_of_mechanical_model}, which are considered in this paper, have the following features \cite{Andrzejewski:2019,Bowden:2012}:
\begin{itemize}
    \item The considered plant is composed of rigid body elements called `links' which are interconnected by particular `joints'.
    \item Movement is considered in $q$ dimensions.
    \item The linear motion of the plant is strictly connected with a proper state variable and it is not constrained.
    \item The angular motion of the plant is strictly connected with a proper state variable and it is constrained to the set $[-180 ~,~ 180] ~ [^\circ]$. 
    \item The movement between plant joints or between plant and surface can be considered as either frictional or not smooth.
\end{itemize}
Moreover, the following assumption is formulated for control system design purposes:
\begin{assumption} \label{as:assumption:one}
Actuator system is located only at the base joint, i.e. the one, which is attached to the either stationary surface or plant moving element, e.g., a  cart or wheels.
\end{assumption}

It is well--known, due to the fact that linear control algorithm can operate in a given operating (equilibrium) point of the non--linear process, it is needed to derive a new, affine (quasi linear) model of the considered system. In general, this model is an approximation of the non-linear vector field in any point $P(\bm{x}_{\mathrm{e}},\bm{u}_{\mathrm{e}})$, which transverse an integral curve assigned to the particular state initial condition $\bm{x}_0$. It is possible to distinguish at least two approaches to devise of linear model \cite{Slotine:1991}. In this work, the approach introduced in \cite{Texteira:1999} has been used. This methodology is based on solving a quadratic optimisation problem. Thereby, problem of applying of incremental variables, which are used in the second main approach basing on the Taylor series expansion, can be omitted and centre of coordinate system is not relocated (redeployed from zero to operating point). Hence, this approach is very convenient due to need of stabilising the considered SIMO class of mechanical dynamical system in one of the equilibrium points. 

\subsection{Affine form of linearised model} \label{subsec:problem_form}

Taking into account that a number of elements of vectors $\bm{\Theta}(t)$ and $\dot{\bm{\Theta}}(t)$ is equal to $n \triangleq 2q \in \mathbb{N}_+$, the state space vector $\bm{x}(t) \in \mathbb{X}$, where $ \mathbb{X} \subset \mathbb{R}^{n}$, can be defined as follows:
\begin{equation} \label{eq:state_vector}
    \bm{x}(t) \triangleq \begin{bmatrix} \bm{\Theta}^{\mathrm{T}}(t) & \dot{\bm{\Theta}}^{\mathrm{T}}(t) \end{bmatrix}^{\mathrm{T}}, ~
    \bm{x_0} \triangleq \begin{bmatrix} \bm{\Theta}_0^{\mathrm{T}} & \dot{\bm{\Theta}}_0^{\mathrm{T}} \end{bmatrix}^{\mathrm{T}}.
\end{equation}
Therefore, considering \eqref{eq:state_vector}, the model \eqref{eq:general_form_of_mechanical_model} yields \cite{Khalil:2002,Isidori:1995}:
\begin{equation} \label{eq:nonlinear_affine_form}
    \begin{cases}
        \dot{\bm{x}}(t) &= \bm{F}(\bm{x}(t)) + \bm{G}(\bm{x}(t))u(t)\\
        \bm{x}(t_0) &= \bm{x}_0
    \end{cases},
\end{equation}
where $\bm{F}(\bm{x}(t))$ is an affine (drift) component of non--linear dynamics and $\bm{G}(\bm{x}(t))$ is a non--linear component associated with control input $u(t)$. 


Taking into account, a particular selection of state variables ensures to the following variant of non--linear affine form \eqref{eq:nonlinear_affine_form}:  
\begin{equation}\label{eq:mechanical_nonlinear_form}
    \begin{cases}
        \dot{x}_1(t) &= x_{q+1}(t)\\
        &\vdots \\
        \dot{x}_{q}(t) &= x_{n}(t)\\
        \dot{x}_{q+1}(t) &= -\bm{C}_1(\bm{x}(t)) \begin{bmatrix} x_{q+1}(t) & \dots & x_{n}(t) \end{bmatrix}^{\mathrm{T}} \\ &- G_1 \left(\begin{bmatrix} x_{1}(t) & \dots & x_{q}(t) \end{bmatrix}^{\mathrm{T}} \right) \\ &+ Q_1\left(\begin{bmatrix} x_{1}(t) & \dots & x_{q}(t) \end{bmatrix}^{\mathrm{T}} \right)u(t) \\
        &\vdots \\
        \dot{x}_{n}(t) &= -\bm{C}_{q}(\bm{x}(t)) \begin{bmatrix} x_{q+1}(t) & \dots & x_{n}(t) \end{bmatrix}^{\mathrm{T}} \\ &- G_{q} \left(\begin{bmatrix} x_{1}(t) & \dots & x_{q}(t) \end{bmatrix}^{\mathrm{T}} \right) \\ &+ Q_{q}\left(\begin{bmatrix} x_{1}(t) & \dots & x_{q}(t) \end{bmatrix}^{\mathrm{T}} \right)u(t)
    \end{cases}.
\end{equation}

Regarding to general form from \eqref{eq:nonlinear_affine_form}, for the first $q$ state equations of \eqref{eq:mechanical_nonlinear_form}, the particular components are equal to: $F_{j}(\bm{x}(t)) = x_{q+j}(t)$ and $G_{j}({\bm{x}(t)}) = 0$. In turn, for the rest of state equations of \eqref{eq:mechanical_nonlinear_form}, the particular components are equal to: $F_{q+j}(\bm{x}(t)) = -\bm{C}_{j}(\bm{x}(t)) \begin{bmatrix} x_{q+1}(t) & \dots & x_{n}(t) \end{bmatrix}^{\mathrm{T}} - G_{j} \left(\begin{bmatrix} x_{1}(t) & \dots & x_{q}(t) \end{bmatrix}^{\mathrm{T}} \right)$ and $G_{q+j}({\bm{x}(t)}) = Q_{j}\left(\begin{bmatrix} x_{1}(t) & \dots & x_{q}(t) \end{bmatrix}^{\mathrm{T}} \right)$, where $j = \overline{1,q}$. It is worth noting that \eqref{eq:mechanical_nonlinear_form} can be decomposed to the linear part and non--linear affine part. 

The linear approximation of \eqref{eq:mechanical_nonlinear_form} can be derived as follows. For any equilibrium point $\bm{x}_{\mathrm{e}} = \begin{bmatrix} x_{\mathrm{e}_1} \ldots x_{\mathrm{e}_n} \end{bmatrix}^{\mathrm{T}} \in \bm{\mathcal{X}}_{\mathrm{e}}$ of the non--linear system \eqref{eq:mechanical_nonlinear_form}, where $\bm{\mathcal{X}}_{\mathrm{e}}$ is a set of all possible equilibrium points such that $\bm{\mathcal{X}}_{\mathrm{e}} = \left\{ \bm{x}_{\mathrm{e}} \in \mathbb{X} \colon x_{\mathrm{e}_j} = 0, ~x_{\mathrm{e}_{q+j}} \neq 0 \right\}$, which is an component of the operating point $P(\bm{x}_{\mathrm{e}},u_{\mathrm{e}})$, where always $u_{\mathrm{e}} = 0$, matrices $\bm{A} = \begin{bmatrix} \bm{a}_{1} & \dots & \bm{a}_{n} \end{bmatrix}^{\mathrm{T}} \in \mathbb{R}^{n \times n}$ and $\bm{B} \in \mathbb{R}^{n \times 1}$ are derived by applying the following formulas \cite{Tatjewski:2007,Texteira:1999}: 
\begin{equation}\label{eq:B_matrix_derivation}
    \bm{B} = \left.\begin{bmatrix} \bm{G}(\bm{x}(t)) \end{bmatrix} \right|_{\bm{x}_{\mathrm{e}}},
\end{equation}
\begin{equation}\label{eq:A_matrix_derivation}
    \begin{split}
        \bm{a}_{\mathrm{i}} &= \left.\begin{bmatrix} \nabla F_{\mathrm{i}}(\bm{x}(t)) \end{bmatrix} \right|_{\bm{x}_{\mathrm{e}}} \\ &+ \dfrac{\left.\begin{bmatrix} F_{i}(\bm{x}(t)) \end{bmatrix} \right|_{\bm{x}_{\mathrm{e}}} - \bm{x}_{\mathrm{e}}^{\mathrm{T}} \left.\begin{bmatrix} \nabla F_{i}(\bm{x}(t)) \end{bmatrix} \right|_{\bm{x}_{\mathrm{e}}}}{||\bm{x}_{\mathrm{e}}||_2}\bm{x}_{\mathrm{e}}
    \end{split},
\end{equation}
where: $\nabla(\cdot)$ denotes gradient of the function; $||\cdot||_2$ is an Euclidean norm of the vector; $i = \overline{1,n}$. 

Thus, using \eqref{eq:B_matrix_derivation} and \eqref{eq:A_matrix_derivation},  the following linear model of the non--linear mechanical affine system \eqref{eq:mechanical_nonlinear_form} can be obtained:

\begin{equation}\label{eq:general_form_of_linear_model}
    \begin{split}
        \begin{cases}
            \dot{\bm{x}}(t) &= \underbrace{\begin{bmatrix} \bm{N} \\ \bm{A}_1 \end{bmatrix}}_{\bm{A}} \bm{x}(t) + \underbrace{\begin{bmatrix} \bm{0}^{q \times 1} \\ \bm{B}_1 \end{bmatrix}}_{\bm{B}} u(t) \\
            \bm{c}(t) &= \underbrace{\begin{bmatrix} \bm{I}^{q \times q} & \bm{0}^{q \times q} \end{bmatrix}}_{\bm{E}} \bm{x}(t) \\
            \bm{x}(t_0) &= \bm{x}_0
        \end{cases},
    \end{split}
\end{equation}
where: $\bm{A}_1 \in \mathbb{R}^{q \times n}$, $\bm{N} = \begin{bmatrix} \bm{0}^{q \times q} & \bm{I}^{q \times q} \end{bmatrix} \in \mathbb{R}^{q \times n}$ are relevant parts of $\bm{A}$ matrix; $\bm{B}_1 \in \mathbb{R}^{q \times 1}$ is a part of $\bm{B}$ matrix; $\bm{I}^{q \times q}$ is an identity matrix of $q \times q$ size; $\bm{0}^{ (\cdot) \times (\cdot)}$ is a zero matrix of an adequate size; $\bm{E} \in \mathbb{R}^{q \times n}$ denotes an output matrix. 

Clearly, taking into account that the first $q$ state equations of \eqref{eq:mechanical_nonlinear_form} are not directly influenced by control input (signal) $u(t)$ and their dynamical character is strictly linear, the appropriate part of matrix $\bm{B}$ equals $\bm{0}^{q \times 1}$. Therefore, matrix $\bm{B}$ of linearised system \eqref{eq:general_form_of_linear_model} must be equal to:
\begin{equation}\label{eq:proof_of_B_matrix}
    \begin{split}
        \bm{B} &= \left.\begin{bmatrix} \bm{G}(\bm{x}(t)) \end{bmatrix} \right|_{\bm{x}_{\mathrm{e}}} = \begin{bmatrix} \bm{0}^{q \times 1} \\ \bm{B}_1 \end{bmatrix}
    \end{split},
\end{equation}
where $\bm{B}_1 =  \left.\begin{bmatrix} \bm{Q}\left(\begin{bmatrix} x_{1}(t) & \dots & x_{q}(t) \end{bmatrix}^{\mathrm{T}} \right) \end{bmatrix} \right|_{\bm{x}_{\mathrm{e}}}$. 

The similar considerations for matrix $\bm{A}$ are as follows. For the first $q$ state equations of \eqref{eq:mechanical_nonlinear_form}, first $q$ elements $\bm{a}_{i}$ of matrix $\bm{A}$ are equal to the particular base vectors $\bm{e}_{q+j} \in \mathbb{R}^{n}$ belong to the Euclidean vector space equipped with a Cartesian coordinate system. Hence, for every pair of $i$ and $j$, which fulfil $i=j$ the following holds:
\begin{equation}\label{eq:proof_of_the_matrix_A_1}
    \begin{split}
            \bm{a}_{i} &= \bm{e}_{q+j} + \dfrac{\left.\begin{bmatrix} F_{j}(\bm{x}(t)) \end{bmatrix} \right|_{\bm{x}_{\mathrm{e}}} - \bm{x}_{\mathrm{e}}^{\mathrm{T}}\bm{e}_{q+j}}{\bm{x}_{\mathrm{e}}^{\mathrm{T}}\bm{x}_{\mathrm{e}}}\bm{x}_{\mathrm{e}} \\ &= \bm{e}_{q+j} + \dfrac{x_{\mathrm{e}_{q+j}} - x_{\mathrm{e}_{q+j}}}{\bm{x}_{\mathrm{e}}^{\mathrm{T}}\bm{x}_{\mathrm{e}}}\bm{x}_{\mathrm{e}} = \bm{e}_{q+j}
    \end{split},
\end{equation}
where $x_{\mathrm{e}_{q+j}}$ is the $q+j$-th element of the $\bm{x}_{\mathrm{e}}$ vector. 

Therefore, it is certain that the `upper' part of the matrix $\bm{A}$ is equal to $\bm{N} = \begin{bmatrix} \bm{0}^{q \times q} & \bm{I}^{q \times q} \end{bmatrix}$. For another $q$ state variables \eqref{eq:A_matrix_derivation} is used explicitly and leading to $\bm{A}_1 = \begin{bmatrix} \bm{a}_{q+1} & \dots & \bm{a}_{n} \end{bmatrix}^{\mathrm{T}}$. Finally, obtained matrices $\bm{N}$ and $\bm{A}_1$ are aggregated and the matrix $\bm{A}$ of \eqref{eq:general_form_of_linear_model} is ensured.

\begin{remark} \label{remark:one}
It is worth noting that the above--presented linear approximation of \eqref{eq:mechanical_nonlinear_form} leads to the same form of $\bm{A}$ and $\bm{B}$ matrices as the Taylor series expansion. However, above method needs putting some general assumptions. If for any equilibrium point $||\bm{x}_{\mathrm{e}}||_2 = 0$, $\bm{x}_{\mathrm{e}}$ has to be transformed to the new $\bm{x}_{\mathrm{e}} \triangleq \bm{x}_{\mathrm{e}} + \bm{\varepsilon}_{\mathrm{e}}$, where $\bm{\varepsilon}_{\mathrm{e}} \in \mathbb{R}^{n}$ is a vector with Euclidean norm: $||\bm{\varepsilon}_{\mathrm{e}}||_2 \rightarrow 0$. This transformation provides to $||\bm{x}_{\mathrm{e}}||_2 \neq 0$ and its justification is straightforward. 
\end{remark}

\begin{remark} \label{remark:two}
Due to the fact that the model \eqref{eq:general_form_of_linear_model} is derived for control system design purposes, it is obvious that the pair $(\bm{A},\bm{B})$ has to be controllable \cite{Astrom:2008}. Taking into account, that considered class of systems, has always the same structure of mathematical linear model \eqref{eq:general_form_of_linear_model}, Kalman controllability matrix is a square matrix denoted by $\bm{M}_{\mathrm{c}} \in \mathbb{R}^{n \times n}$.
\end{remark}

It should be noticed that for some similar class of systems to \eqref{eq:general_form_of_linear_model}, pair $(\bm{A},\bm{B})$ is structurally controllable, if certain particular conditions are fulfilled \cite{Bowden:2012}. These conditions are ensured by the assumption~\ref{as:assumption:one}. Clearly, due to fact that matrix $\bm{A}_1$ has non zero elements, using inductive understanding, it is easy to shown that the system \eqref{eq:general_form_of_linear_model} is structurally controllable.

\section{Synthesis of the control system} \label{sec:control_synthesis}

\begin{figure}[!b]
\centering
\includegraphics[width=2.5in]{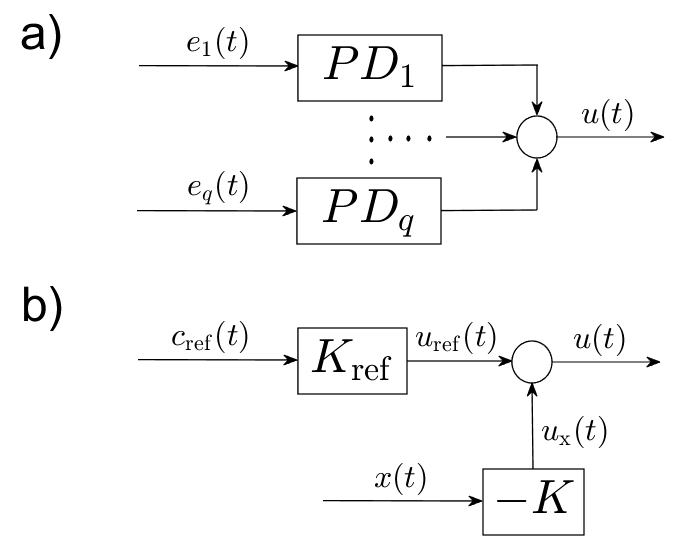}
\caption{The general structures of: a) PD controllers, b) SFR and FFR.}
\label{PD_structure}
\end{figure}

Taking into account that linear model of system \eqref{eq:general_form_of_linear_model} can be used for synthesis of the controller based on the state feedback, according to the topic of this work, it is needed to prove that exist equivalence between structure of the PID type controller and above--mentioned regulator state space. More specifically, the possibility of selection of $q$ PD controllers settings via the State Feedback Regulator (SFR) and Feed Forward Regulator (FFR) is prooved in this section. The general structure of SFR and FFR and the proposed equivalent structure of $q$ PD controllers are presented in Fig.~\ref{PD_structure}.

The aggregated control law of SFR and FFR regulators is as follows \cite{Astrom:2008}:
\begin{equation}\label{eq:aggregated_control_law}
    \begin{split}
        u(t) &\triangleq u_{\mathrm{x}}(t) + u_{\mathrm{ref}}(t); \\
        u_{\mathrm{x}}(t) &\triangleq -\bm{K}\bm{x}(t); ~
        u_{\mathrm{ref}}(t) \triangleq \bm{K}_{\mathrm{ref}} \bm{c}_{\mathrm{ref}}(t),
    \end{split}
\end{equation}
where: $u_{\mathrm{x}}(t)$, $u_{\mathrm{ref}}(t)$ are the SFR`s and FFR`s control signals; $\bm{K} \in \mathbb{R}^{1 \times n}$ is the SFR`s gain matrix; $\bm{K}_{\mathrm{ref}} \in \mathbb{R}^{1 \times q}$ is the FFR`s gain matrix; $\bm{c}_{\mathrm{ref}}(t) \in \mathbb{R}^{q}$ is a vector of the reference signals. The control law \eqref{eq:aggregated_control_law} ensures that in the steady state, which in this case of control problem it can be understood as reaching an equilibrium point: $\lim_{t \to \infty} \bm{c}(t) \rightarrow \bm{c}_{\mathrm{ref}}(t)$. 

In turn, the structure of $q$ PD controllers have to generate one dimensional control signal $u(t)$ and the control law can be written as \cite{Astrom:1995}:
\begin{equation}\label{eq:pd_1}
           u(t) = \bm{K}_{\mathrm{p}}\bm{e}(t) + \bm{K}_{\mathrm{d}} \dot{\bm{e}}(t) = \sum_{j=1}^{q} \left[ K_{\mathrm{p}_{j}} e_{j}(t) + K_{\mathrm{d}_{j}} \dot{e}_{j}(t) \right],
\end{equation}
where: $\bm{e}(t) \triangleq \bm{c}_{\mathrm{ref}}(t) -\bm{c}(t) \in \mathbb{R}^{q}$ is a control error; $\bm{K}_{\mathrm{p}} = \begin{bmatrix} K_{{\mathrm{p}}_1} & \dots & K_{{\mathrm{p}}_q} \end{bmatrix} \in \mathbb{R}^{1 \times q}$ and $\bm{K}_{\mathrm{d}} = \begin{bmatrix} K_{\mathrm{d}_1} & \dots & K_{\mathrm{d}_q} \end{bmatrix} \in \mathbb{R}^{1 \times q}$ denote the vectors of proportional and derivative gains of $q$ PD controllers, respectively. 

The control error derivative with respect to $t$ is equal to:
\begin{equation}\label{eq:pd_2}
    \begin{split}
        \dot{\bm{e}}(t) = \dot{\bm{c}}_{\mathrm{ref}}(t) - \dot{\bm{c}}(t)= \dot{\bm{c}}_{\mathrm{ref}}(t) - \bm{E}\dot{\bm{x}}(t).
    \end{split}
\end{equation}

Taking into account, that reference trajectories are constant over time, the vector $\dot{\bm{c}}_{\mathrm{ref}}(t)$ equals zero vector. Therefore, by substituting \eqref{eq:pd_2} into \eqref{eq:pd_1}, the control law is given as:

\begin{equation}\label{eq:pd_3}
    \begin{split}
        u(t) &= \bm{K}_{\mathrm{p}}\bm{c}_{\mathrm{ref}}(t) - \bm{K}_{\mathrm{p}}\bm{E}\bm{x}(t) - \bm{K}_{\mathrm{d}}\bm{E}\dot{\bm{x}}(t) \\
        &= \bm{K}_{\mathrm{p}}\bm{c}_{\mathrm{ref}}(t) - \begin{bmatrix} \bm{K}_{\mathrm{p}} & \bm{K}_{\mathrm{d}} \end{bmatrix} \bm{E} \begin{bmatrix} \bm{x}^{\mathrm{T}}(t) & \dot{\bm{x}}^{\mathrm{T}}(t) \end{bmatrix}^{\mathrm{T}}.
    \end{split}
\end{equation}

Taking into account the definition of state variable vector and also general form of linear model \eqref{eq:general_form_of_linear_model}, it is obvious that $\bm{E}\bm{x}(t) = \begin{bmatrix} x_{1}(t) & \dots & x_{q}(t) \end{bmatrix}^{\mathrm{T}}$ and $\bm{E}\dot{\bm{x}}(t) = \begin{bmatrix} x_{q+1}(t) & \dots & x_{n}(t) \end{bmatrix}^{\mathrm{T}}$. Thus, \eqref{eq:pd_3} can be rewritten as:
\begin{equation}\label{eq:pd_4}
    u(t) = \bm{K}_{\mathrm{p}}\bm{c}_{\mathrm{ref}}(t) - \begin{bmatrix} \bm{K}_{\mathrm{p}} & \bm{K}_{\mathrm{d}} \end{bmatrix} \bm{x}(t).
\end{equation}

By comparing control laws \eqref{eq:pd_4} and \eqref{eq:aggregated_control_law}, the settings of PD controllers structure is obtained from the following equations:
\begin{equation}\label{eq:pd_5}
    \begin{split}
        \bm{K}_{\mathrm{ref}} = \bm{K}_{\mathrm{p}}, ~\bm{K} = \begin{bmatrix} \bm{K}_{\mathrm{p}} & \bm{K}_{\mathrm{d}} \end{bmatrix}.
    \end{split}
\end{equation}

Hence, the equivalence between  the controller based on the state feedback and the PID type controller has been proved.


\section{Case study - a two--wheeled balancing robot} \label{sec:case_study}

A balancing robot is a single--axle, twin--track vehicle which centre of mass is above the axis of the road wheels (see Fig. \ref{fig:robot}). It is an example of autonomous mobile constructions, which belong to the class of considered SIMO systems.

\subsection{Model of the two--wheeled balancing robot} \label{subsec:model_case_study}
A non--linear state--space mathematical model of a two--wheeled balancing robot which was derived in \cite{Laddach:2019} is given as follows:
\begin{equation}\label{eq:modelNierozwiklany}
    \begin{cases}
        \dot{x}_{\mathrm{1}}(t) = x_{\mathrm{3}}(t), \dot{x}_{\mathrm{2}}(t) = x_{\mathrm{4}}(t),\\
    (I_{\mathrm{n}}+m_{\mathrm{n}} l^2)\dot{x}_{\mathrm{3}}(t) = -m_{\mathrm{n}}l\dot{x}_{\mathrm{4}}(t)\cos(x_{\mathrm{1}}(t))
    + \frac{2 k k_{\mathrm{m}} k_{\mathrm{e}}}{R r}x_{\mathrm{4}}(t)\\ 
    -\frac{2 k_{\mathrm{m}} k_{\mathrm{e}}}{R}x_{\mathrm{3}}(t)
    + m_{\mathrm{n}} g l \sin(x_{\mathrm{1}}(t)) 
    - \frac{2 k_{\mathrm{m}}}{R}u(t),\\
    (\frac{2I_{\mathrm{k}}}{r^2} + 2m_{\mathrm{k}} + m_{\mathrm{n}})\dot{x}_{\mathrm{4}}(t) = \frac{2 k k_{\mathrm{m}} k_{\mathrm{e}}}{R r}x_{\mathrm{3}}(t) - \frac{2k^2 k_{\mathrm{m}} k_{\mathrm{e}}}{R r^2}x_{\mathrm{4}}(t) \\
    - m_{\mathrm{n}}l\dot{x}_{\mathrm{3}}(t)\cos(x_{\mathrm{1}}(t)) + m_{\mathrm{n}} l x_{\mathrm{3}}^2(t)\sin(x_{\mathrm{1}}(t))+\frac{2 k k_{\mathrm{m}}}{R r}u(t),
    \end{cases}
\end{equation}
where: $x_1(t) ~[^\circ]$ is the angular displacement (tilt); $x_2(t) ~[\mathrm{m}]$ denotes the linear displacement; $x_3(t) ~[^\circ/\mathrm{s}]$ stands for the angular velocity; $x_4(t) ~[\mathrm{m/s}]$ signifies the linear velocity; $u(t) ~[\mathrm{V}]$ is the voltage applied to the DC motor. Using the methodology presented in \cite{Andrzejewski:2019}, model \eqref{eq:modelNierozwiklany} can be rewritten into the ODE form resembling to \eqref{eq:nonlinear_affine_form}:
\begin{equation}\label{eq:cogtitive_model_of_robot}
    \begin{cases}
        \dot{x}_1(t) &= x_3(t) \\
        \dot{x}_2(t) &= x_4(t) \\
        \dot{x}_3(t) &= \dfrac{H_3(\bm{x}(t))}{M(\bm{x}(t))} + \dfrac{P_3(\bm{x}(t))}{M(\bm{x}(t))}u(t)  \\
        \dot{x}_4(t) &= \dfrac{H_4(\bm{x}(t))}{M(\bm{x}(t))} + \dfrac{P_4(\bm{x}(t))}{M(\bm{x}(t))}u(t) \\
        \bm{x}(t_0) &= \bm{x}_0, \\
        \bm{c}(t) &= \begin{bmatrix} 1 & 0 & 0 & 0 \\ 0 & 1 & 0 & 0 \end{bmatrix} \bm{x}(t).
    \end{cases}
\end{equation}
 The elements $H_3(\bm{x}(t))$, $H_4(\bm{x}(t))$, $P_3(\bm{x}(t))$, $P_4(\bm{x}(t))$ and $\forall t ~ M(\bm{x}(t)) \neq 0$ are defined as:
\begin{equation}\label{eq:model_components}
    \begin{split}
        H_3(\bm{x}(t)) &= \alpha_6\alpha_1 + \alpha_7\alpha_3, ~
        H_4(\bm{x}(t)) = \alpha_6\alpha_3 + \alpha_7\alpha_2,\\
        P_3(\bm{x}(t)) &= \alpha_4\alpha_1 + \alpha_5\alpha_3, ~
        P_4(\bm{x}(t)) = \alpha_4\alpha_3 + \alpha_5\alpha_2,\\
        M(\bm{x}(t)) &= \alpha_1 \alpha_2 - \alpha_3^2,
    \end{split}
\end{equation}
where all of the $\alpha_{(\cdot)}$ are as follows:
\begin{equation}\label{eq:model_parameters}
 \begin{split}
    \alpha_1 &= I_{\mathrm{n}} + m_{\mathrm{n}}  l^2, ~
    \alpha_2 = 2\dfrac{I_{\mathrm{k}} }{r^2} + 2m_{\mathrm{k}}  + m_{\mathrm{n}},\\
    \alpha_3 &= -m_{\mathrm{n}}l\cos(x_1(t)), ~
    \alpha_4 = -2\dfrac{k_{\mathrm{m}}}{R}, ~ 
    \alpha_5 = 2\dfrac{kk_{\mathrm{m}}}{Rr},\\
    \alpha_6 &= 2x_4(t)\dfrac{kk_{\mathrm{e}}k_{\mathrm{m}}}{Rr} - 2x_3(t)\dfrac{k_{\mathrm{e}}k_{\mathrm{m}}}{R} + m_{\mathrm{n}}gl\sin(x_1(t)),\\
    \alpha_7 &= 2x_3(t)\dfrac{kk_{\mathrm{e}}k_{\mathrm{m}}}{Rr} - 2x_4(t)\dfrac{k^2k_{\mathrm{e}}k_{\mathrm{m}}}{Rr^2} + m_{\mathrm{n}}l\sin(x_3(t))^2.
 \end{split}
\end{equation}
The parameter values of the model \eqref{eq:model_parameters} are given as:
\begin{itemize}
    \item $I_{\mathrm{n}} = 0.0112 ~ [\mathrm{kg} \cdot \mathrm{m^2}] $ - the moment of inertia of the robot construction,
    \item $I_{\mathrm{k}} = 3.9337 \times 10^{-5} ~ [\mathrm{kg} \cdot \mathrm{m^2}]$ - the moment of inertia of the robot wheel,
    \item $m_{\mathrm{n}} = 1.12 ~ [\mathrm{kg}]$ - the mass of the robot,
    \item $m_{\mathrm{k}} = 0.125 ~ [\mathrm{kg}]$ - the mass of the robot wheel,
    \item $l = 0.1 ~ [\mathrm{m}]$ - the distance to the robot centre of gravity,
    \item $R = 2.1428 ~ [\mathrm{\Omega}]$ - the winding resistance,
    \item $r = 0.045 ~ [\mathrm{m}]$ - the wheel radius,
    \item $k = 34.014 ~ [-]$ - the gear ratio,
    \item $k_{\mathrm{e}} = 68.9655 \times 10^{-4} ~ [\mathrm{V} \cdot \mathrm{s}]$ - the electro--mechanical constant,
    \item $k_{\mathrm{m}} = 14.8850 \times 10^{-2} ~ [\mathrm{V} \cdot \mathrm{s}]$ - the torque constant,
    \item $g = 9.81 ~ [\mathrm{m/s^2}]$ - the gravitational acceleration.
\end{itemize}

The detailed derivation of model \eqref{eq:model_parameters} can be found in \cite{Laddach:2019}.

\subsection{The control problem}

As it has been mentioned above, the aim of the control system of the two--wheeled balancing robot is its stabilisation at the given  equilibrium  point, which equals $\bm{x}_{\mathrm{e}} = \begin{bmatrix} 0 & 0 & 0 & 0 \end{bmatrix}^{\mathrm{T}}$ in the considered case. This control goal is fulfilled using the proposed linear control algorithms. Hence, taking into account remark \ref{remark:one} it is assumed that $\bm{\varepsilon}_{\mathrm{e}} = \bm{1}^{4\times1} \times 10^{-4}$, where $\bm{1}^{(\cdot) \times (\cdot)} $ denotes a matrix with all elements equal one. Therefore, by the use of \eqref{eq:B_matrix_derivation} and \eqref{eq:A_matrix_derivation}, model \eqref{eq:cogtitive_model_of_robot} can be rewritten in general form \eqref{eq:general_form_of_linear_model} with the matrices $\bm{A}$ and $\bm{B}$ given as:
%
\begin{equation}\label{eq:robot_model_AB}
        \begin{split}
            \bm{A} &= \begin{bmatrix} 0 & 0 & 1 & 0 \\ 0 & 0 & 0 & 1 \\ 1.4188 & 5.7939 \times 10^{-7} & -4.3319 & 3274.4 \\ -0.1128 & -6.6786 \times 10^{-12} & 0.8586 & -648.99 \end{bmatrix}, \\ \bm{B} &= \begin{bmatrix} 0 & 0 & -628.4856 & 124.4993
        \end{bmatrix}^{\mathrm{T}}.
    \end{split}
\end{equation}

The Kalman controllability matrix $\bm{M}_{\mathrm{c}} = \begin{bmatrix} \bm{B} & \bm{A}\bm{B} & \bm{A}^{2}\bm{B} & \bm{A}^3\bm{B} \end{bmatrix}$ can be used to show that the pair $(\bm{A},\bm{B})$ is controllable. It is ensured because $\bm{M}_{\mathrm{c}}$ is non--singular matrix due to $\det(\bm{M}_{\mathrm{c}}) = -1.4517 \times 10^{13}$.

In order to design the SFR gain matrix, the optimisation approach -- LQR has been used. Since the analysis of the selection of values in the diagonal matrices  $\bm{Q}_{\mathrm{LQR}} \in \mathbb{R}_{+}^{n \times n}$ and $R_{\mathrm{LQR}} \in \mathbb{R}_{+}$ has not been the subject of the paper, they have been chosen arbitrarily, with the assumption that the first and second state variables are more significant as:
\begin{equation}\label{eq:LQR_matrices}
    \bm{Q}_{\mathrm{LQR}} = \text{diag}(100,100,1,1),~ R_{\mathrm{LQR}} = 1.
\end{equation}

Hence, by solving the linear--quadratic optimisation problem the state feedback gain matrix has been obtained. Moreover, according to \eqref{eq:pd_5} the proportional and derivative gains of the $q$ PD controllers have been ensured:
\begin{equation}\label{eq:gains_of_regualtors}
    \begin{split}
        \bm{K} &= \begin{bmatrix} \bm{K}_{\mathrm{p}} & \bm{K}_{\mathrm{d}} \end{bmatrix} \\ &= \begin{bmatrix} -13.1881 & -10.0 & -9.3717 & -45.1452 \end{bmatrix}.
    \end{split}
\end{equation}

\begin{remark} \label{remark:four}
The discrete control law is needed for the physical implementation of PD controllers. This control law includes a low--pass filter with sampling time equals $T_{\mathrm{s}} = 0.1 ~[\mathrm{s}]$ \cite{Czyzniewski:2019}. Moreover, due to DC voltage saturation, the limitation of the control signal to $ u(t) \in [-12~,~12] ~[\mathrm{V}]$ has been taken into account \cite{Laddach:2019}.
\end{remark}

\subsection{Simulation results}

\begin{figure}[!b]
\centering
\includegraphics[width=2.4in]{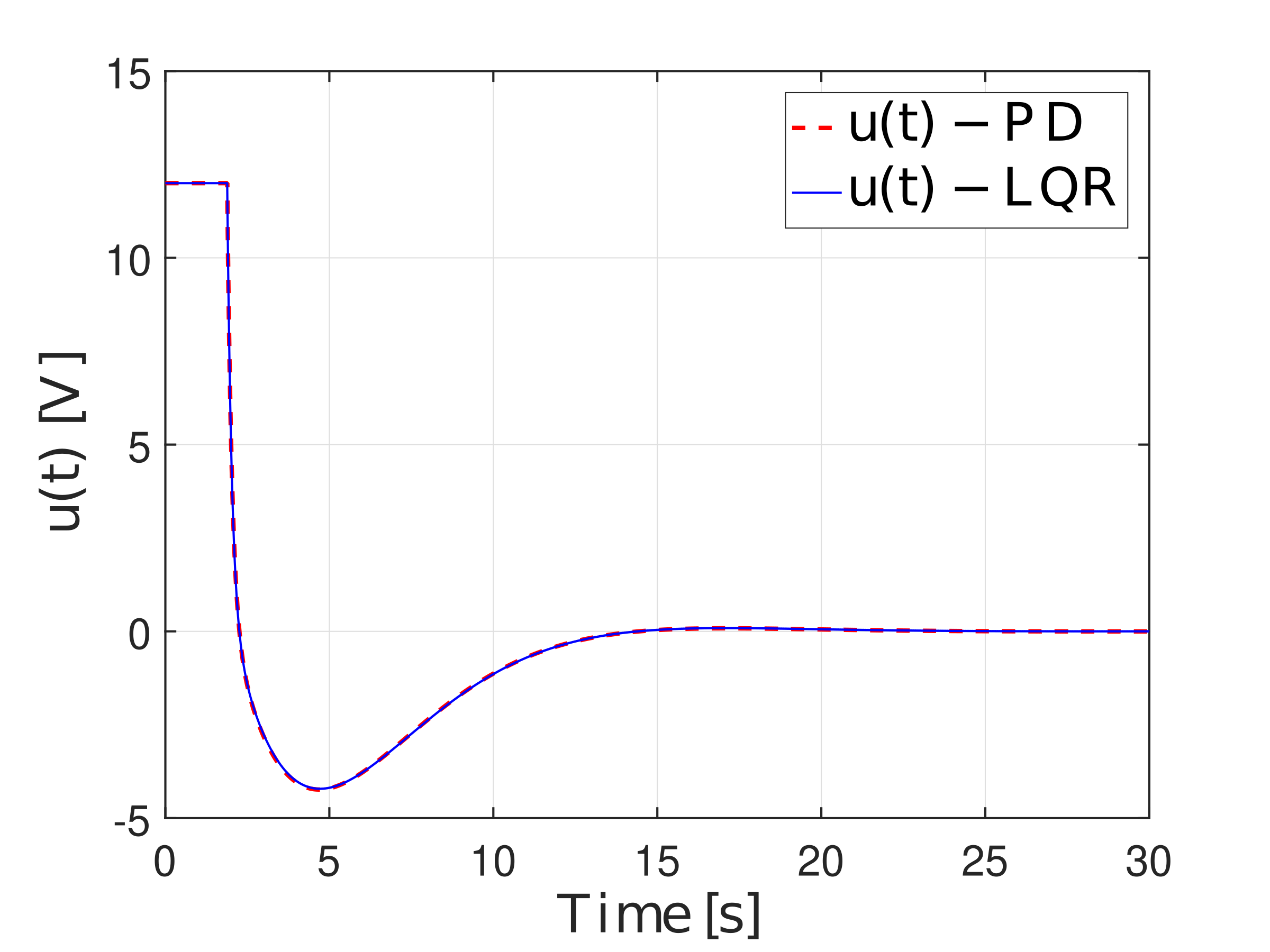}
\caption{Trajectories $u(t)$ -- PD and SFR continuous.}
\label{sim_1}
\end{figure}

\begin{figure}[!t]
\centering
\includegraphics[width=2.4in]{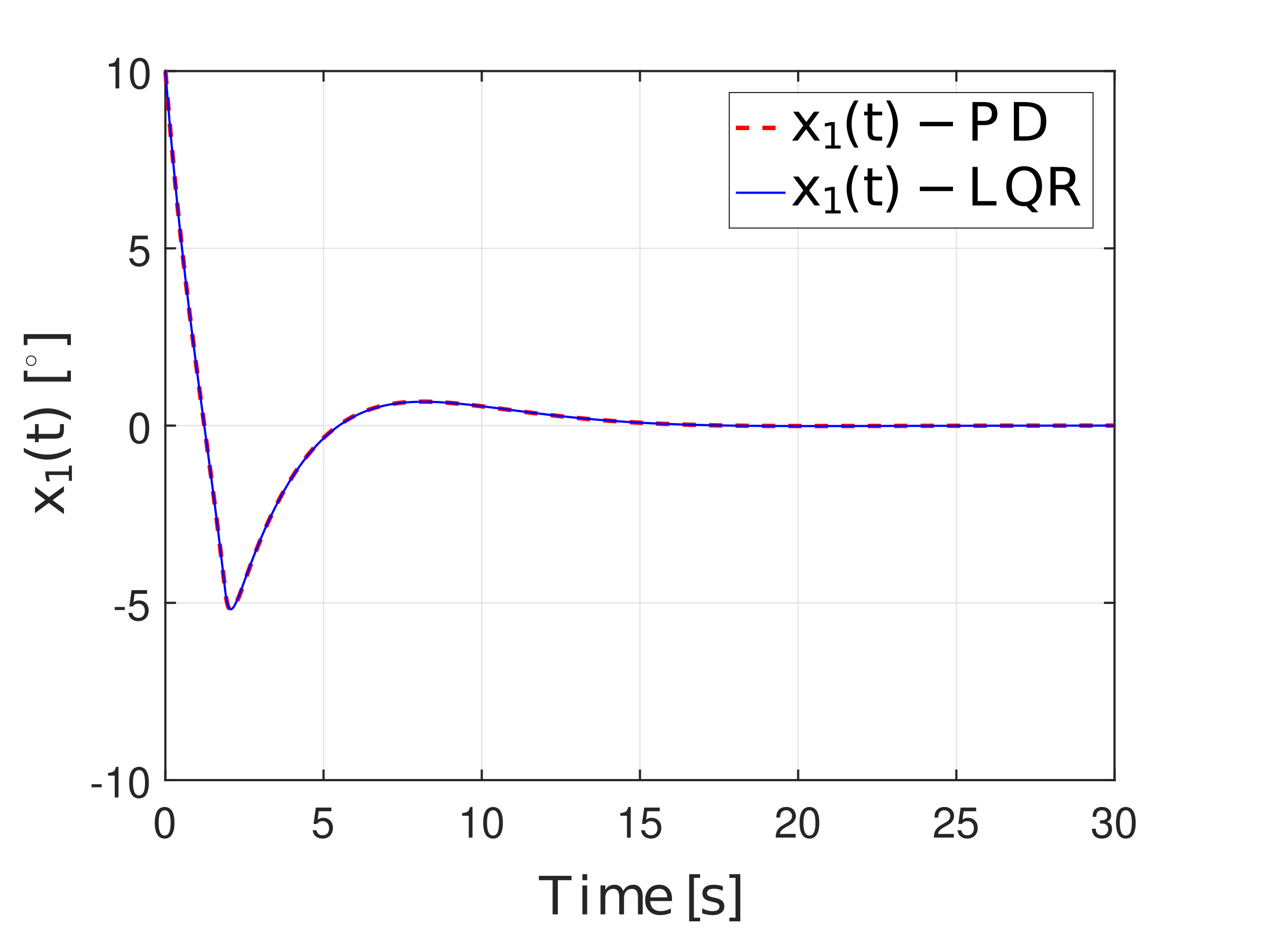}
\caption{Trajectories $x_1(t)$ -- PD and SFR continuous.}
\label{sim_2}
\end{figure}

\begin{figure}[!t]
\centering
\includegraphics[width=2.4in]{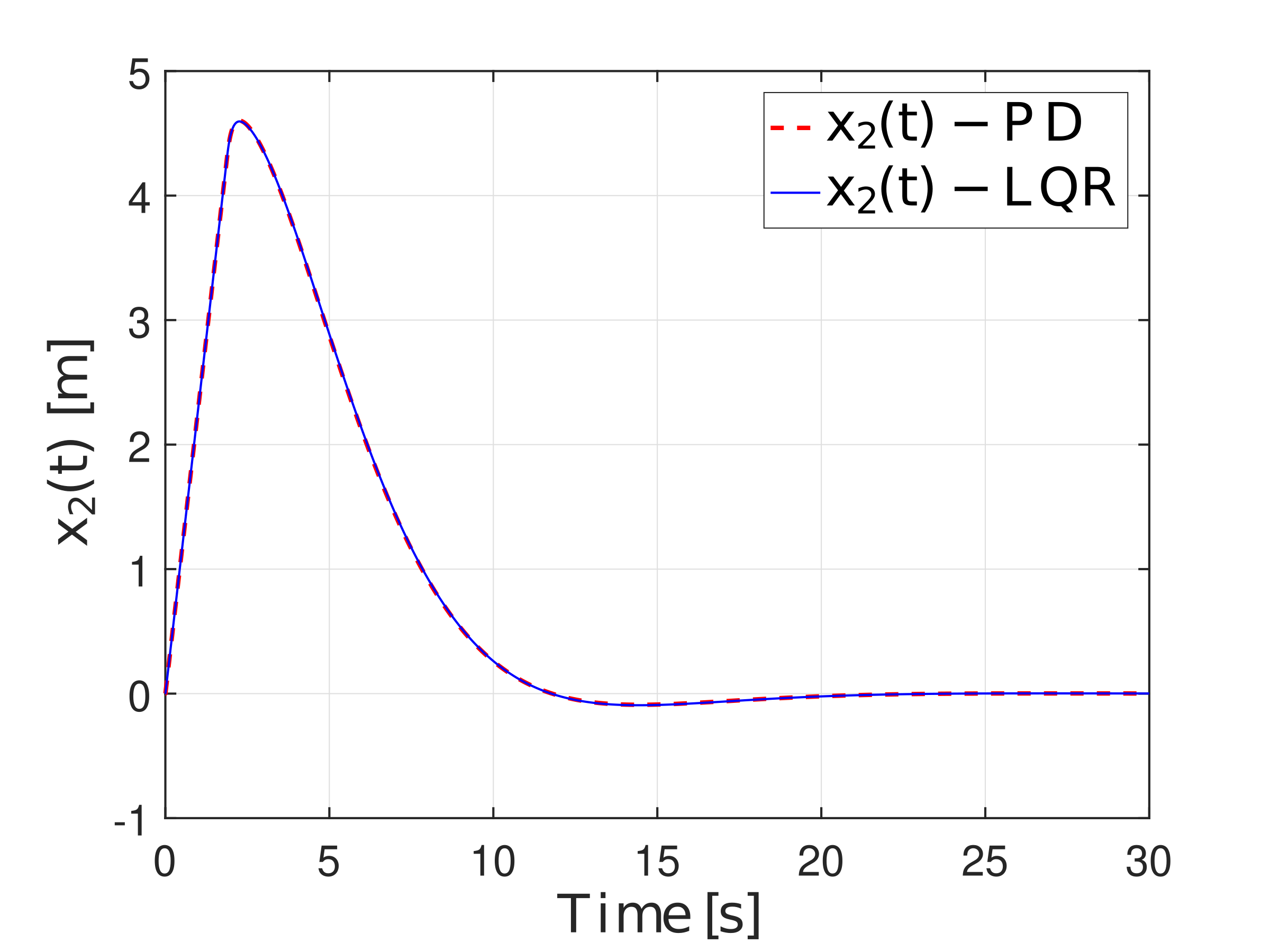}
\caption{Trajectories $x_2(t)$ -- PD and SFR continuous.}
\label{sim_3}
\end{figure}

\begin{figure}[!b]
\centering
\includegraphics[width=2.7in]{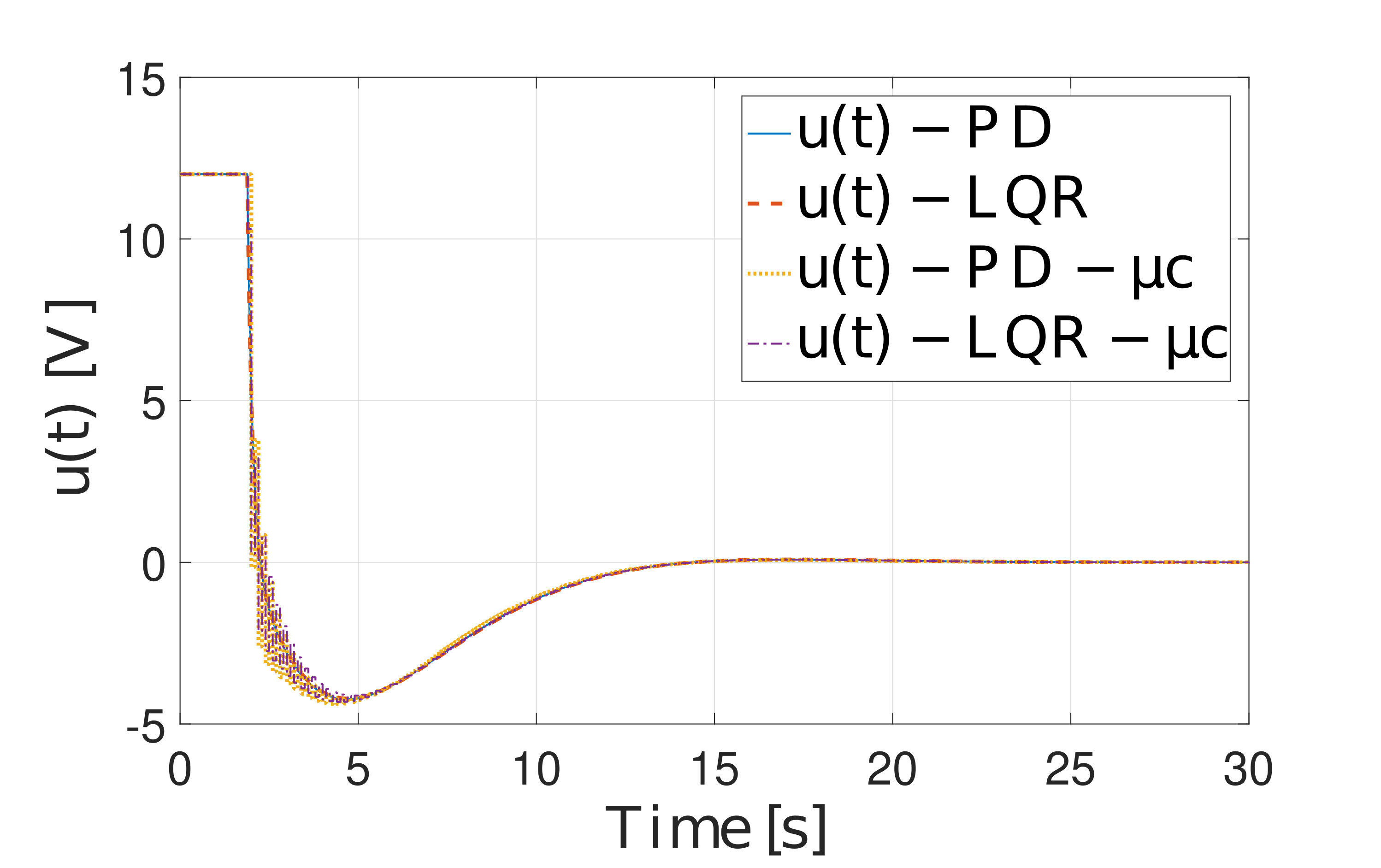}
\caption{Trajectories $u(t)$ -- PD and SFR continuous and discrete.}
\label{sim_4}
\end{figure}

\begin{figure}[!t]
\centering
\includegraphics[width=2.7in]{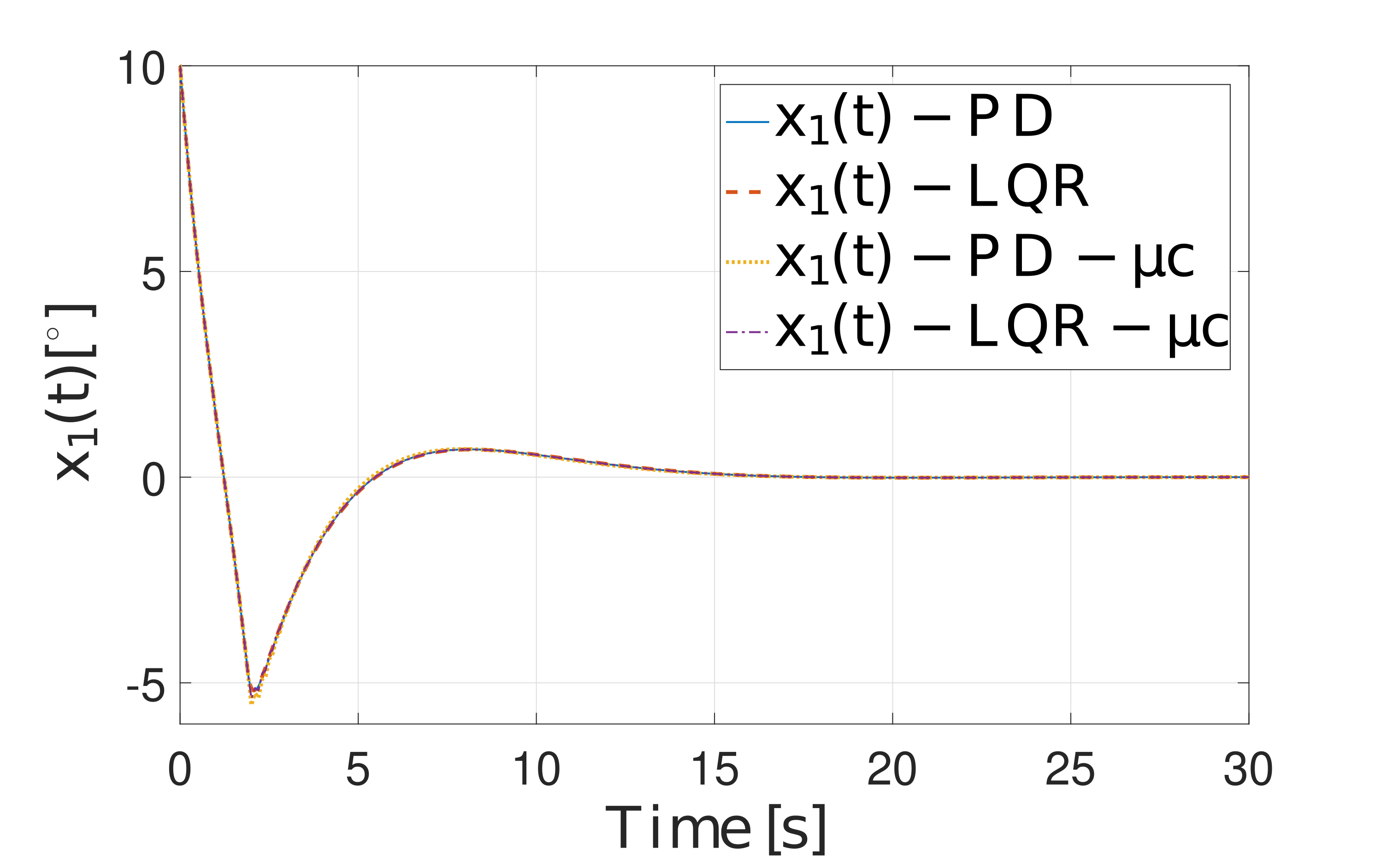}
\caption{Trajectories $x_1(t)$ -- PD and SFR continuous and discrete.}
\label{sim_5}
\end{figure}

\begin{figure}[!t]
\centering
\includegraphics[width=2.7in]{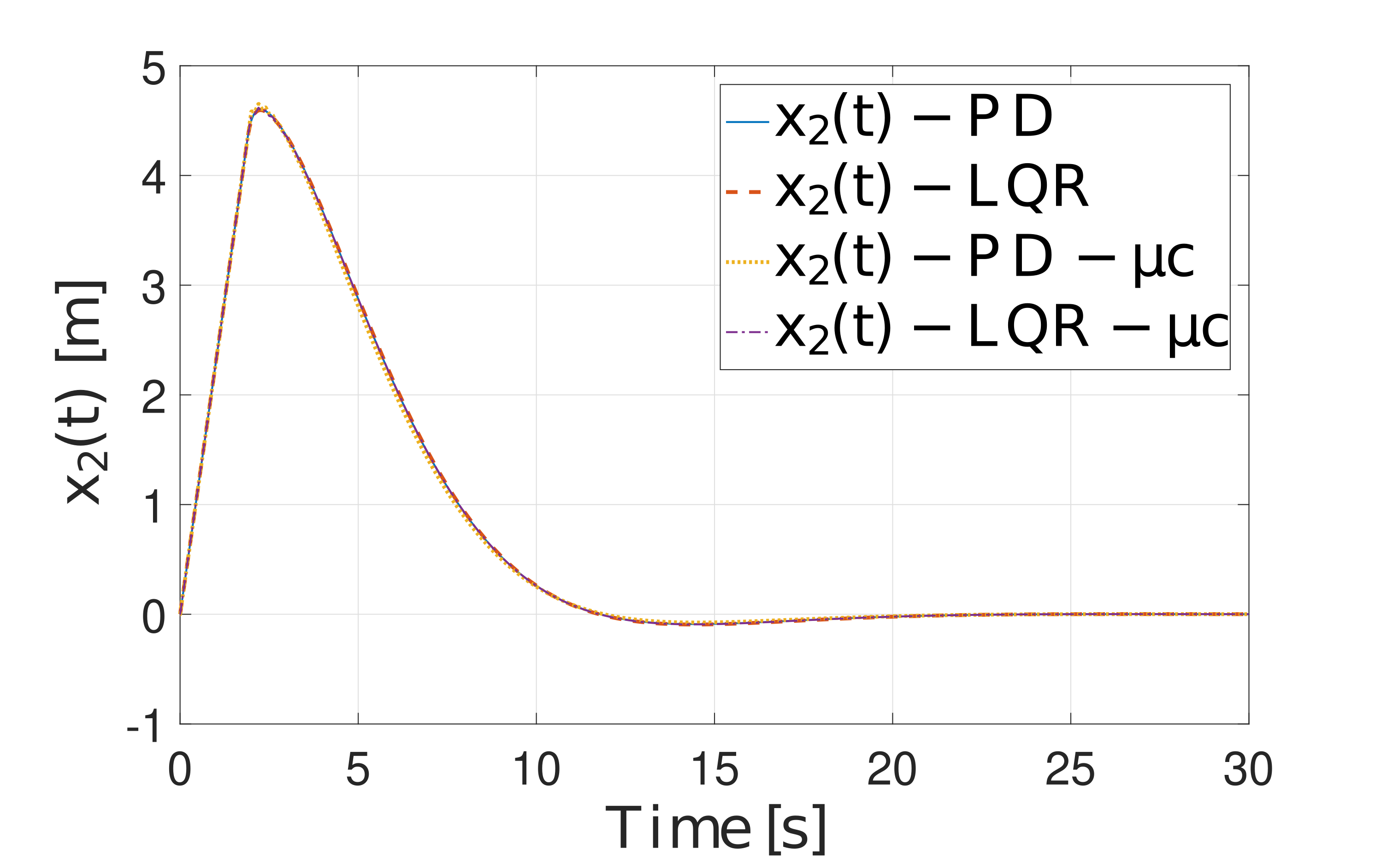}
\caption{Trajectories $x_2(t)$ -- PD and SFR continuous and discrete.}
\label{sim_6}
\end{figure}

The both control systems (see Fig. \ref{PD_structure}) with the set of parameters which is shown in subsection \ref{subsec:model_case_study} have been implemented and validated in Matlab/Simulink environment. The results of representative simulation experiments are presented in Figs.~\ref{sim_1}--\ref{sim_6}. These results present the performance of stabilisation regulators (SFR and PD) and it has been qualitatively assessed. The initial conditions have been arbitrarily selected as $\bm{x}_0 = \begin{bmatrix} 10 & 0 & 0 & 0 \end{bmatrix}^{\mathrm{T}}$, which represent 10–degrees tilt the robot from the considered equilibrium point $\bm{x}_{\mathrm{e}}$.

First, the continuous realisation of PD and SFR controllers has been investigated. The comparisons between trajectories of $u(t)$, $x_1(t)$ and $x_2(t)$ are shown in Figs.~\ref{sim_1}--\ref{sim_3}, respectively.

As it can be noticed in the both cases, the signal $u(t)$ applied to the robot causes its angular displacement, which also involves linear displacement of the robot. After about 15 seconds the robot has been stabilised at the $\bm{x}_{\mathrm{e}}$. This naturally results in $u(t)$ reaching zero. Moreover, the trajectories of the control signal generated by the PD and SFR are only slightly different, which directly proves the possibility of effective PD tuning using the gains of the state feedback regulator.

Next, in Figs.~\ref{sim_4}--\ref{sim_6} the trajectories from a simulated discrete environment (simulated discrete hardware implementation) have been added. 

The results presented in Figs.~\ref{sim_4}--\ref{sim_6} shows that the discrete realisations of the PD and SFR controllers generates significantly different control signals with the comparison to the continuous ones. However, properly chosen sample time $T_{\mathrm{s}}$ ensures stability and a good performance of the entire system.

\section{Conclusion} \label{sec:conclusion}

In this paper, the problem of selection of the PID type controller settings for the class of SIMO mechanical dynamical systems has been investigated. In particular, the possibility of transforming optimal settings of the linear--quadratic regulator (LQR) into the settings of the PD controller has been given. The equivalence of both structures has been shown at the design stage for the continuous realisation of the particular regulators. In turn, the discrete algorithms developed for hardware implementation needs the usage of low pass filtering and constrain of the control signal, due to peaking phenomena caused by numerical differentiation of control error. However, as it has been shown in the simulation way, the discrete controller provides insignificantly weaker performance with comparison to the continuous control law. Hence, the selection of State Feedback Regulator (SFR) as an optimal regulator (LQR) ensures that the equivalence PD controller has also optimal proportional and derivative gains.  

\section*{Acknowledgements}
The research work was done in accordance with funding from Polish MEiN under Young Researcher Support Program. The authors wish to express their thanks for support.

\bibliographystyle{IEEEtran}%
\bibliography{ref}%

\end{document}